\documentclass[article,prl,twocolumn]{revtex4}
\usepackage{color}

\usepackage{graphicx}

\begin{document}
\title{Annihilation of Point Defect Pairs in Freely Suspended Liquid-Crystal Films}

\author{Amine Missaoui$^{1\ast}$, Kirsten Harth$^{1\ast}$, Peter Salamon$^2$, Ralf Stannarius$^1$}
	\affiliation{%
		$^1$Institute of Physics, Otto von Guericke University, 39106 Magdeburg, Germany,\\
		$^2$Wigner Research Centre of Physics, Budapest, Hungary
        \footnote{$^\ast${AM and KH contributed equally to this manuscript.}}
	}

\begin{abstract}
We study the annihilation of topological defect pairs in the quasi-twodimensional (2D) geometry of freely suspended smectic films.
This elementary process is at the basis of all models describing the statistics of complex defect patterns.
We prepare pairs with opposite topological charges and retrieve the interaction mechanisms from their trajectories.
The square-root dependence of the defect separation on the time until annihilation and the asymmetry in propagation velocities of the opponents
predicted by theory are confirmed. The importance of defect orientations is demonstrated. Trajectories
are in general curved, depending on the mutual orientations (phase mismatch) of the defects and on the orientation of the pair respective to the far, undisturbed director. The experiments provide the basis for an adaption of the theoretical models to the real complexity of the annihilation.
\end{abstract}

%\flushbottom
\maketitle

\thispagestyle{empty}

%\section*{Introduction}

Topological defects occur in a wide variety of physical systems, for example in soft matter \cite{Poulin1997,Poulin1998,Musevic2006,Brugues2008,Tkalec2011,Guimaraes2013,Musevic2019}, quantum systems \cite{Onsager1949,Weiler2008,Polkovnikov2011}, thin magnetic films \cite{Wachowiak2002,Hertel2006}, superfluid liquids \cite{Ruutu1996,Bauerle1996,Bauerle1996b}, and even cosmology \cite{Volovik1977,Chuang1991}. Often, complex defect patterns are generated after symmetry-breaking phase transitions. Their coarsening dynamics can be
essential for the establishment of the new, ordered state.

Many features of defect dynamics are universal.
Liquid crystals (LCs) were suggested as ideal model systems to study such phenomena \cite{Chuang1991,Bowick1994,Zurek1996,Kibble2013}.
%Results of LC experiments can be compared to similar topological structures e.g. in cosmology \cite{Volovik1977}.
Their defect dynamics can be observed in facile polarizing microscopy experiments, with comparably simple equipment.
The elementary process of pair annihilations of topologically opposite-charged point defects allows to construct scaling solutions for more complex defect patterns.

Nematics form the simplest LC meso\-phase, yet experiments with nematics in sandwich cells are not easy to interpret, because of the influence of cell boundaries and the generally 3D character of the deformations. Smectic C (SmC) freely suspended films \cite{Young} are the quasi-2D analogue of a polar nematic, ideally suited to study topological defect dynamics \cite{Pleiner1988,Svensek2003,Svensek2003E,Radzihovski2015,Muzny1994,Wachs2014,Stannarius2016}.
Here, we explore the role of defect orientations in pairs with opposite topological charges during annihilation.

\begin{figure}[ht!]
	\centering
	\includegraphics[width=0.66\columnwidth]{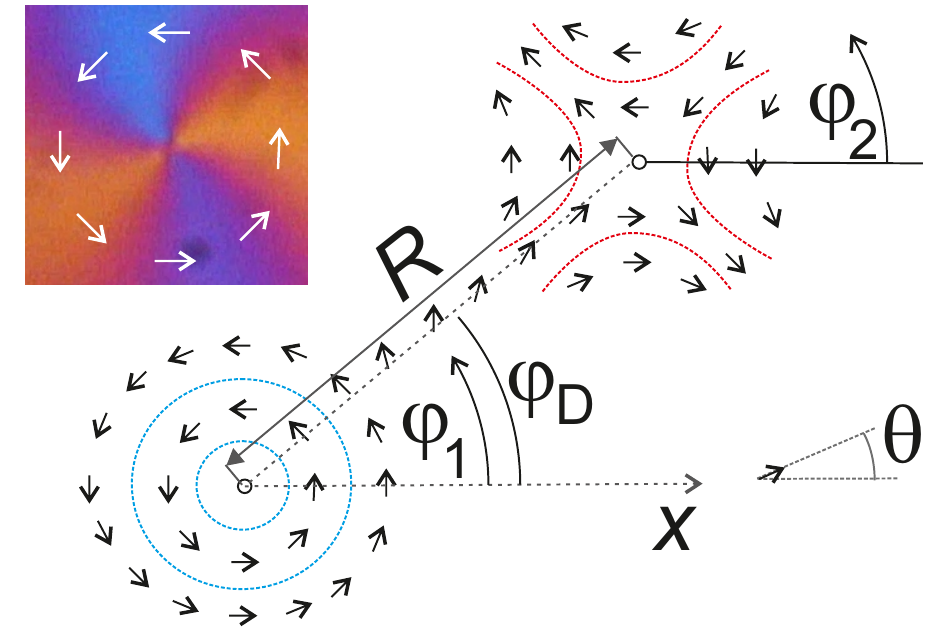}
\caption{\label{fig:1}
Point defects with topological charges $+1$ (bottom left) and $-1$ (top right) in mismatch
($\delta\theta \neq 0$): Along the straight connection of the two cores, the c-director rotates.
$\theta$ is the c-director angle to the $x$ axis, $R$ is the defect separation, $\varphi_{1,2}$ and $\varphi_D$ are explained in the text.
The top-left photo illustrates the relation between c-director (arrows) and color in the polarizing microscope with
crossed polarizers and a diagonal $\lambda$ phase plate.
% The pair is also in misalignment with the far director field.
}
\end{figure}

In SmC, the mesogens have a preferred tilt to the smectic layer normal.
The c-director (projection of the tilt direction onto the film plane) characterizes the local orientation. Its dynamics is well described by continuum models. The 2D character of the problem simplifies modeling and reduces boundary effects to the far-away film edges. Optical textures reflect the local c-director orientations, the types and positions of defects. Yet even though such experiments are apparently quite simple, there are only few reports on defect dynamics in SmC films \cite{Muzny1992,Muzny1994,Pargellis1994,Wachs2014,Stannarius2016}, none about pair annihilation.
The polar c-director can only form defects of integer strengths. It shares this feature with all other systems where vortices of vector fields
are relevant (e.~g.~\cite{Wachowiak2002,Hertel2006}). A scenario of primary interest is the
annihilation of pairs with topological charges $S_{1,2}=\pm1$ (We may set $S_1=+1,S_2=-1$ without loss of generality). The angle $\theta$ of the c-director with the
$x$ axis at positions $\vec r=(x,y)$ near the defect cores $\vec R_i =(x_i,y_i),\  i=\{1,2\}$, is $\theta=\theta_{i}+S_{i}\varphi_i$, where $\theta_i$ are the phases of the defects and $\varphi_i$ are the angles of the relative positions $\vec r-\vec R_i$ with the $x$ axis (see Fig.~\ref{fig:1}).

Models for defect interactions were developed for nematics in one-constant elastic theory
\cite{Dafermos1970,Minoura1996,Kleman2003}, and extended to SmC films \cite{Pleiner1988,Svensek2003,Svensek2003E,Radzihovski2015},
where elastic anisotropy \cite{Svensek2003,Svensek2003E} and dynamic influences \cite{Radzihovski2015} were included.
The simplest model is derived from linear superposition of single-defect equilibrium solutions of the director field \cite{Dafermos1970}.
It assumes that the defects pass quasi-equilibrium states. The force between them, acting along the separation vector $\vec R=\vec R_2-\vec R_1$,
is
$ %\begin{equation}
   -2 \pi K    /   R
$ %\end{equation}
 \cite{Chandrasekhar1986,Kleman2003}.
$K$ is a mean elastic constant, $R=|\vec R|$. The drag forces on the defects are approximately proportional to  $\dot R$ \cite{Chandrasekhar1986}, thus
one obtains % $\dot R \propto R^{-1}$ and thus
a time dependence
$  %\begin{equation}
R = \sqrt{2D_{1}(t_{0}-t)}\label{eq:4},
$  %\end{equation}
where $t_{0}$ is the annihilation time, $D_1$ is a diffusion coefficient that contains $K$, the rotational viscosity $\gamma$, and the Ericksen number
\cite{Chandrasekhar1986}.
The defects approach each other on straight paths.
Because of an asymmetric coupling to backflow, the $+1$ defect is predicted to move considerably faster than the $-1$ opponent
\cite{Svensek2003}. The pair does not annihilate halfway, but closer to the initial $-1$ position.

All these models implicitly assume that the defect orientations match, i.~e. that the c-direc\-tor is constant along the straight connection
between the defect cores.
Vromans and Giomi \cite{Vromans2016} noticed that the so far disregarded mutual orientations are essential features.
Tang and Selinger \cite{Tang2017}, generalized this idea to arbitrary defect strengths.
% The geometry considered in their model is relevant in our experiments.
For conjugated pairs with $S_1=1$ and $S_2=-1$, their equations read
\begin{eqnarray}\label{eq:10}
%\begin{split}
\theta(\vec r)  &=&  \arctan\left(\frac{y-y_{1}}{x-x_{1}}\right) -  \arctan\left(\frac{y-y_{2}}{x-x_{2}}\right)\\
  &+&\frac{\delta\theta}{2}
\left[1+\frac{\log(|\vec{r}-\vec{R_{1}}|)-\log(|\vec{r}-\vec{R_{2}}|)}
{\log(R)-\log(r_{\rm c})}\right]
   +\theta_0 \nonumber
\\
	\delta\theta &=& \theta_2 - \theta_1 - 2 \varphi_D -  \pi ,\,
	\theta_0= \theta_1 + \varphi_D +\pi.\nonumber
%	\end{split}
	\end{eqnarray}
The generalized equations for conjugated defect pairs with charges $S_{1,2}=\pm S$ are found in Appendix A.
Note that the terms $-\pi$ and $+\pi$ in the definitions of $\delta\theta$ and $\theta_0$ arise from the correct choice of the quadrants
of the arctan functions used in Ref.~\cite{Tang2017}.
While these equilibrium distortions are exact, they do not preserve the c-director far from the defects,
$
\theta_{\infty}  = {\delta\theta}/{2}+\theta_0 = ({\theta_2 + \theta_1+\pi})/{2}.
$
In experiments, one usually studies defect dynamics under fixed boundary conditions, thus one needs to rotate these solutions to fix $\theta_\infty$.
We choose $\theta_\infty'=0$ (primed angles in the rotated system) without loss of generality. In the primed system rotated by $-\theta_\infty$,
the equilibrium solutions are given by
\begin{equation}\label{eq:2}
\varphi_D' = -\pi-\theta_1-\frac{\delta\theta}{2},\quad \theta'_1=\theta_1,\quad
 \theta_2' = -\theta_1-\pi . %= \frac{5}{2}\pi.
 \end{equation}
The mismatch remains unchanged, $\delta\theta'=\delta\theta$, and the phase of the $+1$ defect is preserved.
Note that $\varphi'_D$ and $\delta\theta$ are not independent of each other in equilibrium. After a
coordinate transformation that fixes $\theta_\infty'$, it becomes clear that the $-1$ defect always chooses an
equilibrium mismatch {\em angle} in accordance with its {\em position} and vice versa in a given external c-director field, corresponding
to the energetic minimum. This aspect was not noticed in Ref. \cite{Tang2017} where director field in infinity was disregarded.
%In the second part of their paper, the authors performed numerical simulations which produced curved paths of mismatched defect pairs.

These models presuppose elastic isotropy, i.~e. equal elastic constants for c-director splay, $K_S$, and bend, $K_B$.
This may be used as a reasonable first approximation for our experiments (where $K_S\approx 2.2K_B$), except in the very vicinity of the $+1$ defect.
When $K_B$ is lower than $K_S$, which is the predominant situation,
all $+1$ defects adopt tangential orientation ($\theta_1=\pm \pi/2$) in equilibrium. Such a pinning of the alignment angle at a $+1$ vortex is
typical for other systems, too, e.~g. in magnetic thin films. The deformation near the core is pure bend,
and the c-director is pinned near the core. Even small differences  $K_B - K_S$ suffice to fix the phase.
This was demonstrated in previous experiments \cite{Stannarius2016} which revealed some limitations of the classical models in SmC films.

This pinning is the only specific aspect that we need to add in the description of our system: we set $\theta_1 = \pi/2$. For $\theta_1=-\pi/2$, all
conclusions will be the same except that one has to change the sign of the c-director.
The $-1$ defects are less affected by elastic anisotropy, their director field is only slightly modified, and the phase $\theta_2$ only rotates the defect.
During the annihilation process, a mismatched pair ($\delta\theta\neq 0$) either has to move to an appropriate angle $\varphi_D'$,
or the $-1$ defect has to rotate (change $\theta_2'$), or both.

In our experiments, we demonstrate the interrelation of the two important orientation parameters, the phase mismatch $\delta\theta$ of the pair
and the misalignment $\delta\phi=\varphi_D'+\pi+\theta_1$ relative to the far c-director. The definition of $\delta\phi$ is chosen such that it
reaches zero when the defect orientations match, $\delta\theta=0$. This is a reasonable choice, as $\theta_1+(\varphi'_D+\pi)$ is exactly the c-director
angle near the $+1$ core on the side opposing the conjugate defect. The pair is considered aligned when this angle equals $\theta_\infty$, i. e. when the $-1$ defect is opposing the direction where $\theta=0$ near the $+1$ defect.

We performed experiments with a room temperature smectic C mixture of 50:50 wt.\% 5-n-octyl-2-[4-(n-hexyloxy)phenyl]-pyrimidine and  5-n-decyl-2-[4-(n-octyloxy) phenyl]-pyrimidine.
Defects were created by touching the homogeneously oriented free-standing film with a hair tip. In 24 experiments, we obtained isolated defect pairs.
Initial separations $R$ were of the order of 200 $\mu$m. The initial alignment angle could not be controlled, it was determined
a posteriori from the video images. The defect pairs were
observed in a microscope under crossed polarizers with a diagonal $\lambda$ wave plate (550 nm, slow axis from top right to bottom left). Video frame rates were 30 fps, in a few experiments 50 fps.
The defect positions are easily localized in the images, with an accuracy of approximately 1~$\mu$m.
The c-director orientations are retrieved from texture colors, the accuracy is  $\approx 10^\circ$.
The relation between c-director $\vec c$  and optical texture is indicated in Fig. \ref{fig:1}.
We cannot determine the sign of $\vec c$ with our observation technique, thus we choose a given sense of direction and use this assignment consistently for all experiments.
This has no consequences for the evaluation of data, the equations are independent of the sign of
the tilt azimuth. It is possible that all arrows in the images actually represent $-\vec c$.

\begin{figure}[htbp!]
	\centering
	\includegraphics[width=\columnwidth]{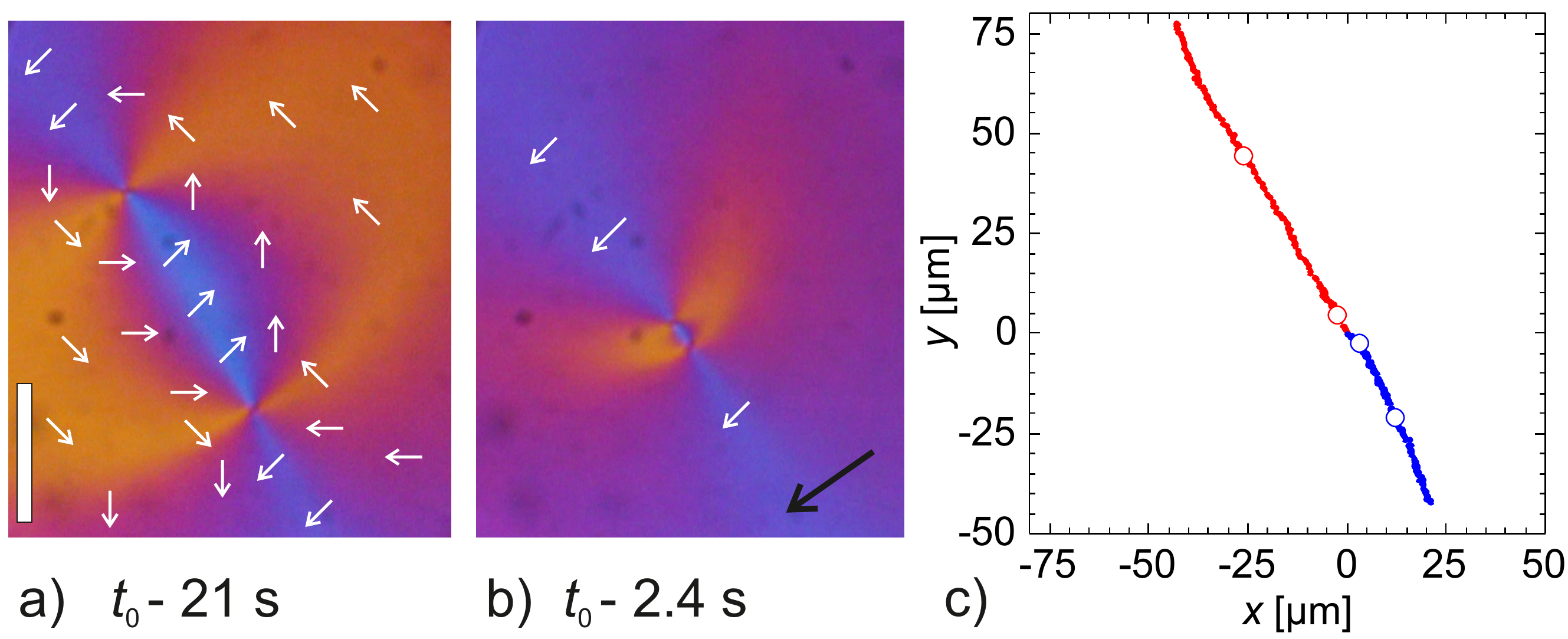}
\caption{\label{fig:2}
(a,b) Matching ($\delta\theta\approx 0$) and aligned ($\delta\phi\approx 0$)  defect pair on the way to annihilation. White arrows sketch the c-director. The $+1$ defect (top) is approximately 1.8 times faster than the $-1$ defect. The black arrow in (b) indicates the outer director field. (c) Defect trajectories respective to the annihilation point. The circles mark the positions in frames (a) and (b). The white bar is 50 $\mu$m.
}
\end{figure}

Figure \ref{fig:2} shows an initially matching ($\delta\theta=0$) and aligned ($\delta\phi \approx 0$) defect pair.
Such conditions (within approximately 10$^\circ$ accuracy) were achieved coincidentally in 5 experiments.
Eq.~(\ref{eq:2}) predicts that when the orientations match, the misalignment with the director will be $\pi/2$. This agrees with our observations.
Immediately before annihilation, we found $\varphi_D'\rightarrow -\pi-\theta_1=-3\pi/2$ and $\delta\phi\rightarrow 0$ in all experiments except one. When this relation is coincidentally fulfilled
already after preparation, then $\delta\theta$ and $\delta\phi$ remain constant and the defects follow straight paths.
The $+1$ defect is much faster than the $-1$, in qualitative agreement with theory \cite{Toth2002,Svensek2003}.
\begin{figure}[htbp!]
	\centering
	\includegraphics[width=\columnwidth]{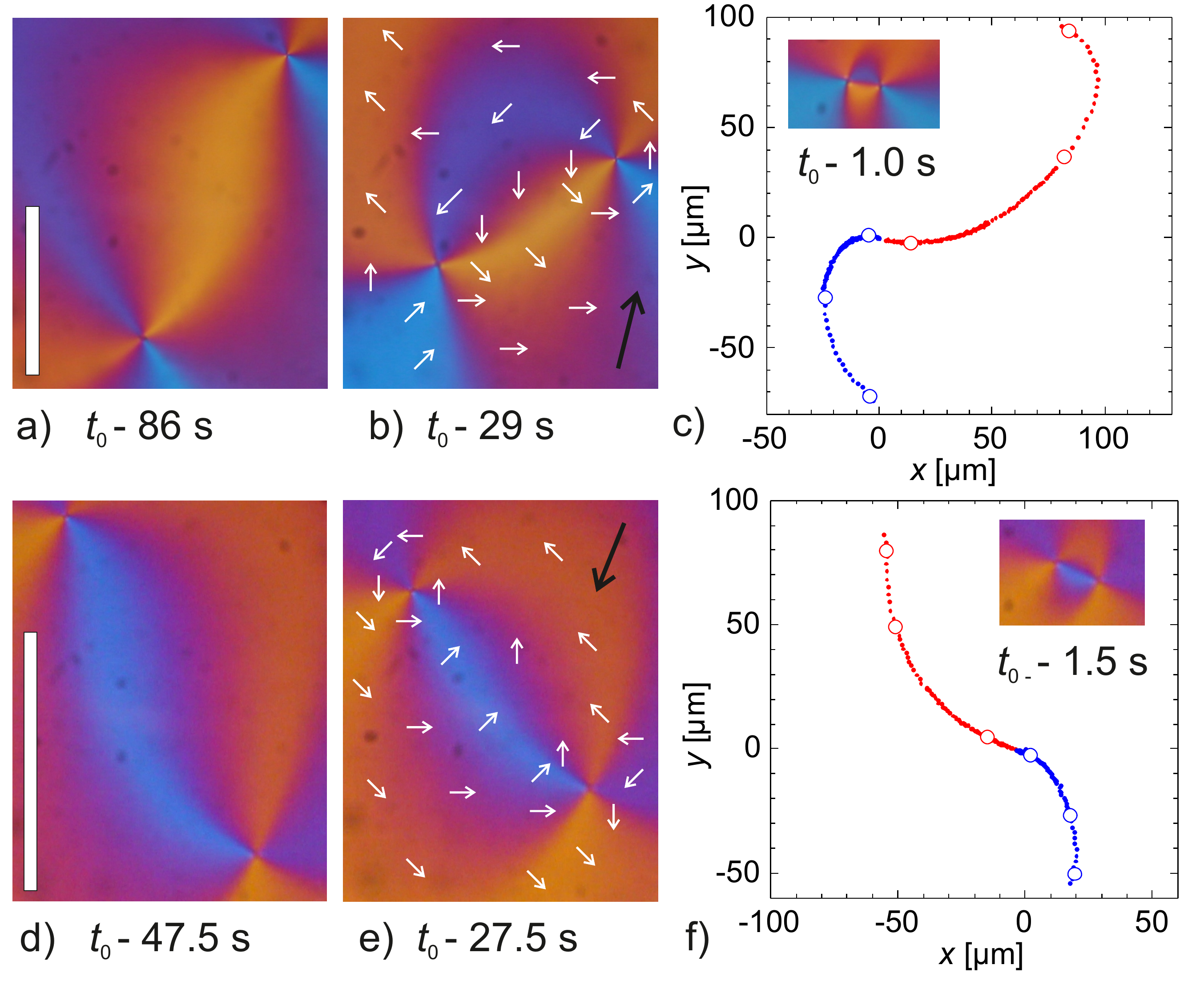}
\caption{\label{fig:3}
(a,b) Mismatched (initial $\delta\theta\approx 65^\circ$) and misaligned (initial $\delta\phi\approx -36^\circ$)
defect pair on the way to annihilation. White arrows sketch the c-director. The $+1$ defect (top) is $\approx 1.8$
times faster than the $-1$ defect. The black arrow indicates the outer director field.
(c) Trajectories respective to the annihilation point. Circles mark the defect positions in frames (a-c).
(d,e) Mismatched (initial $\delta\theta\approx -50^\circ$) and misaligned (initial $\delta\phi\approx 79^\circ$) pair.
(f) Trajectories, circles mark the defect positions in frames (d-f). The velocity ratio is $\approx 1.7$.
The white bars represent 100~$\mu$m.
}
\end{figure}

\begin{figure*}[htbp!]
	\centering
	\includegraphics[width= 0.98\textwidth]{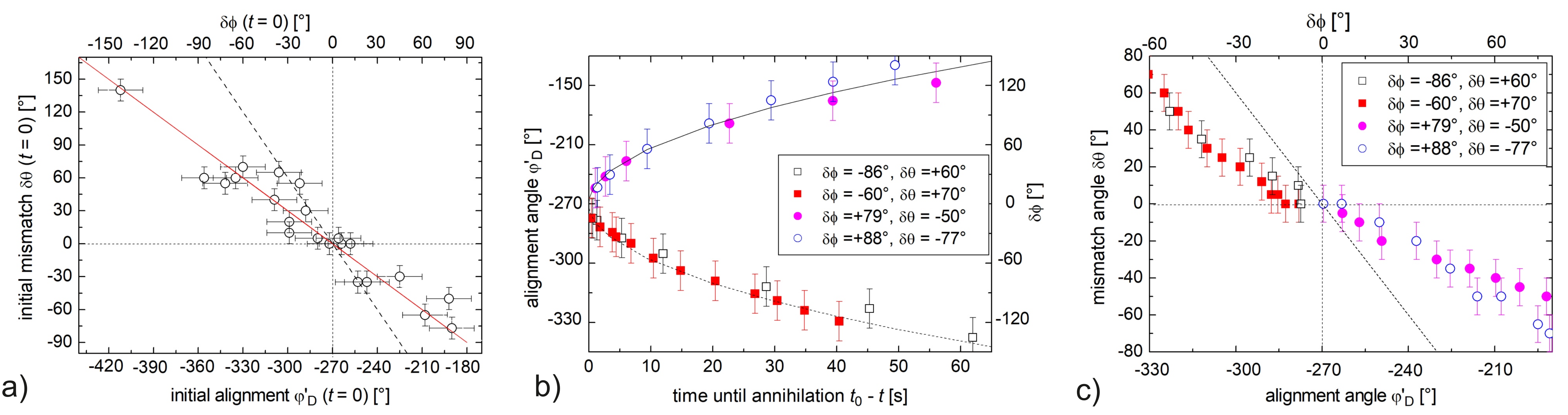}
\caption{\label{fig:a}
a) Time dependence of the pair alignment angle $\varphi'_D$ for four selected defects. The solid and dashed lines represent
$\delta\phi=\pm 9.0 ^\circ\sqrt{t_0-t}/\sqrt{\rm s} $.
b) Mismatch $\delta\theta$ vs. alignment $\varphi_D'$, the solid line represents $\delta\theta=-\delta\phi$.
Even though in individual experiments there are deviations of up to 20$^\circ$, it is obvious that the two angles are correlated.
Eq.~(\ref{eq:2}) predicts a linear dependence with slope  $-2$ (dashed line) for the quasi-equi\-librium configurations.
c) Mismatch angles vs. alignment angles during the approach of exemplary defect pairs with S and mirrored-S trajectories. The dashed line
represents Eq.~(\ref{eq:2})
}
\end{figure*}

The other 19 experiments randomly produced defect pairs with either positive or negative misalignment $\delta\phi$. Figure \ref{fig:3}(a-c) shows a typical example with
initially negative $\delta\phi$ and positive $\delta\theta$. The c-director changes along the defect connection line, and the texture
adopts an S-shape. The trajectory is no more straight, it has mirrored-S-shape.
The $+1$ defect is still faster than the $-1$ defect. This trajectory of Fig.~\ref{fig:3}(c) is typical for all similar initial conditions.
The angle $\varphi_D'$ changes during the approach, finally reaching $90^\circ$. Immediately before annihilation
the defects are both orientation matched, $\delta\phi\rightarrow 0$, and aligned, $\delta\phi\rightarrow 0$.
This holds for all defect pairs independent of the initial angles.
The case of initially negative $\delta\theta$ and positive $\delta\phi$ produces the opposite curvature,
as shown in Fig. \ref{fig:3}(d-f). The texture
forms a mirrored S  between the defects, the trajectory has S-shape. For all defect pairs,
the misalignment $\delta\phi$ follows a square-root dependence on the time to annihilation (Fig.~\ref{fig:a}a).
%
%\begin{figure}[htbp!]
%	\centering
%	\includegraphics[width=0.8\columnwidth]{timedep4.png}
%\caption{\label{fig:5}
%Time dependence of the pair alignment angle $\varphi'_D$ for four selected defects. The solid and dashed lines represent
%$\delta\phi=\pm 9.0 ^\circ\sqrt{t_0-t}/\sqrt{\rm s} $.
%
%}
%\end{figure}

For the test of the predictions of Ref. \cite{Tang2017}, it would be informative to observe other initial combinations of $\delta\theta$ and $\delta\phi$, e.~g., both with the same sign. In fact, this was never achieved in our experiment. All initial combinations are plotted in Fig.~\ref{fig:a}b).
There is an obvious correlation, irrespective of the deviations that are most probably caused by global distortions of the film orientation when the defects are created. Equation~(\ref{eq:2}) suggests a relation $\delta\theta=-2\delta\phi$ in the quasi-equilibrium states
(elastic energy minima at fixed defect positions), which is in clear contrast to the experiment.
%\begin{figure}[htbp!]
%	\centering
%	\includegraphics[width=0.85\columnwidth]{misal-mism4.png}
%\caption{\label{fig:4}
%Mismatch $\delta\theta$ vs. alignment $\varphi_D'$, the solid line represents $\delta\theta=-\delta\phi$.
%Even though in individual experiments there are deviations of up to 20$^\circ$, it is obvious that the two angles are correlated.
%Eq.~(\ref{eq:2}) predicts a linear dependence with slope  $-2$ (dashed line) for the quasi-equi\-librium configurations.
%}
%\end{figure}

The preparation technique produces random alignment angles $\varphi_D'$, but it seems unlikely that touching
the films creates correlated combinations of $\delta\theta$ and $\delta\phi$ (even though this cannot be strictly excluded). We cannot see the first
few seconds after defect formation for technical reasons, but we conclude that the defects rearrange within this short period after creation, irrespective of the initial conditions,
into an orientation that fulfills the condition $\delta\theta=-\delta\phi$, at least approximately. It is reasonable to assume that the rotation of the
$-1$ defect
in place can proceed much faster than a spatial displacement of the same defect on a circle around the opponent. The reason why there is a factor
of nearly 2 between the calculated equilibrium states \cite{Tang2017} and the experimental observation at least for large $|\delta\phi|$ is not clear.
A dynamic solution of the differential equations for the c-director might elucidate this.

We conclude that (1) the system establishes certain combinations of $\delta\theta$ and $\delta\phi$ spontaneously, irrespective of
the details of the defect creation, and (2) either the experimental system does not develop through quasi-equilibrium states
on the way to annihilation, or the equilibrium states are different from those predicted from one-constant elastic theory.
A qualitative estimation of the effects of elastic anisotropy $K_S\neq K_B$ on the equilibrium alignment is given in Appendix B.
%
%\begin{figure}[htbp!]
%	\centering
%	\includegraphics[width=0.85\columnwidth]{indi4.png}
%\caption{\label{fig:6}
%Mismatch angles vs. alignment angles during the approach of exemplary defect pairs with S and mirrored-S trajectories. The dashed line
%represents the quasi-equilibrium states, Eq.~(\ref{eq:2})
%
%}
%\end{figure}

The `initial' angles in Figs. \ref{fig:3} and \ref{fig:a}a are in a way arbitrary, they were taken after the defect preparation as soon as the microscope was
focussed onto the film surface in the spot where the defects appeared. This takes a few seconds, differing between individual runs.
Initial fast reorientations of the c-director field are not accessible. After these transients, the relations between misalignment and
mismatch obviously remain fixed until annihilation, where both $\delta\phi$ and $\delta\theta$ reach zero.
As a test of this invariance, Fig. \ref{fig:a}c) shows four typical experiments, two with S-shaped and two with mirrored S-shaped trajectories.
The two angles $\delta\theta$ and $\delta\phi$ remain strongly correlated, roughly proportional to each other, although there is a slight tendency
of $\delta\theta$ to relax faster than $\delta\phi$. Again, quasi-equilibrium configurations would correspond to $\delta\theta=-2\delta\phi$. This is the consequence of the fixed $\theta_1$, a property that our system shares, e. g., with thin permalloy films \cite{Wachowiak2002,Hertel2006}.

With respect to the model \cite{Tang2017}, our hypothesis is that the equilibrium configurations for $K_S=K_B$ are not appropriate quantitative descriptions
of the states the annihilating defect pair passes on its approach. The torque predicted from the quasi-equilibrium configurations in Ref.~\cite{Tang2017} is probably causing an initial reorientation of the $-1$ defect, which is too fast to be observed in our experiment.
The relation $\delta\theta =-\delta\phi$ observed in our study is presumably a balance
between the torque exerted by the $+1$ opponent and the torque exerted by the far c-director field. The latter is obviously not modeled
in Tang's study \cite{Tang2017} in a way appropriate to describe our experiment with fixed boundary conditions.

The relaxation on curved trajectories, which at first glance looks very similar to the numerical simulations in Refs.~\cite{Vromans2016,Tang2017},
appears to be primarily caused by the torque of the external director on the topological dipole.
One can consider this as an analogue, with some peculiarities, of an electrical dipole in a uniform external field. The electrical dipole tends to
align parallel to the external field to minimize its energy. Likewise, the conjugated topological defect pair can
optimize its elastic energy, no matter what the mismatch angle is, by choosing a proper orientation $\varphi_D'$ respective to the uniform far c-director $\vec c_0$. If one defines the `topological dipole', for instance, as a unit vector $\vec p=(\vec{R_2}-\vec{R_1})/R$ pointing from the $+1$ to the $-1$ defect, the energetic minimum is defined by the angles in Eq.~(\ref{eq:2}). In our experiments,
where $\theta_1=\pi/2$, the topological dipole aims to align with an angle $\varphi_D'=(-\pi-\delta\theta)/{2}$ respective to $\vec c_0$. We note that the situation of a fixed $\theta_1=\pm \pi/2$ is quite common, not only for smectic films, in particular in systems where the divergence of the vector field is suppressed or inhibited (vortex flow of an incompressible liquid or splay of a magnetic or electric polarization). In the general case of arbitrary integer or half-integer charges $\pm S$, one finds equilibrium orientations of the topological dipole defined by the vector from the $+S$ to the $-S$ core
\begin{equation}
\label{eq:S9}
\varphi'_D = - \frac{\delta\theta}{2S} - \frac{1}{S} \theta'_1 - \pi.
\end{equation}
(see Appendix A).
The relaxation of $\delta\phi$ follows a similar square-root time dependence as the defect separation $R$. This produces the observed curved trajectories as seen in Fig.~\ref{fig:2}. In half-integer pairs, both defects can rotate and the situation is more complex (see Appendix A), but the combination of the two torques of the conjugate defect and the far field, respectively, remains effective.

We have demonstrated that the shape of trajectories of mutually annihilating topological defect pairs is equally influenced by their relative orientations (mismatch) and the surrounding vector field (misalignment), both being of equal importance. They are not independent of each other. We suspect that this is a common feature of topological defect pairs in many other 2D systems. During approach, misalignment and mismatch decay to zero, depending on the time until annihilation by a square-root law. The classical theories \cite{Dafermos1970,Minoura1996,Kleman2003,Svensek2003} remain valid for the special case of aligned, matching pairs. The c-director configurations during the approach are qualitatively similar to quasi-equilibrium states determined from the solutions of the Laplace equation \cite{Tang2017}, but differ quantitatively. This may be caused in our films by flow-coupling
\cite{Svensek2003}, elastic anisotropy \cite{Svensek2003,Brugues2008}, or the finite film area. An explanation of this discrepancy requires further studies.

\section*{Appendix A: Director fields around conjugated point defects}
\label{sec1}
We consider a vector field (c-director) or director field (n-director) which is characterized by a unit vector that is oriented at an
angle $\theta(\vec r)$ respective to the $x$ axis in a two-dimensional plane $(x,y)$. Further, we assume that the equilibrium configurations
are defined by solutions of the Laplace equation. For the liquid crystal systems considered here, this means that we assume that the
elastic constants for bend, $K_B$, and splay, $K_S$, are equal.

For general pairs of conjugated defects with $S_1=S$ and $S_2=-S$, the equations given by Tang and Selinger \cite{Tang2017} for the
solutions of the Laplace equation read
\begin{eqnarray}\label{eq:S1}
%\begin{split}
\theta(\vec r)  &=& S\arctan\left(\frac{y-y_{1}}{x-x_{1}}\right) - S\arctan\left(\frac{y-y_{2}}{x-x_{2}}\right) \nonumber\\
  &+&\frac{\delta\theta}{2}
\left[1+\frac{\log(|\vec{r}-\vec{R_{1}}|)-\log(|\vec{r}-\vec{R_{2}}|)}
{\log(R)-\log(r_{\rm c})}\right]
   +\theta_0 , \nonumber \\
	 \delta\theta &=& \theta_2 - \theta_1 - 2S\varphi_D - S\pi ,\nonumber \\
	\theta_0 &=& \theta_1 +S(\varphi_D +\pi).
%	\end{split}
	\end{eqnarray}

The terms $-S\pi$ and $+S\pi$ in the definitions of $\delta\theta$ and $\theta_0$ arise from the correct explicit choice of the quadrants
of the arctan functions used in Ref.~\cite{Tang2017}.
The angle $\varphi_D$ defines the orientation of the connection vector from the core of the $+1$ defect to the core of the $-1$ defect
respective to the coordinate axis $x$.
The c-director field far from the two defects approaches the uniform value $\theta_\infty= \delta\theta/2+\theta_0$, i.~e.
\begin{equation}
\theta_\infty = \frac{\theta_1+\theta_2+S\pi}{2},
\end{equation}
which depends upon the phases of the two defects, which are in general not constant during their mutual approach and annihilation.
In experiments, the far director field is usually fixed, and independent of the defect configurations. Therefore, it is useful
for an interpretation of the equilibrium configurations to consider them with respect to the far c-director orientation.
We rotate the coordinate system such that the infinitely far c-director is along $x'$, i. e. $\theta'_\infty=0$. Thus,
the new equations for the c-director field near the two defect cores are transformed to
$
\theta' = \theta'_1 + S\varphi'
$.
With the transformations $\theta' = \theta-\theta_\infty$, $\varphi'=\varphi-\theta_\infty$, the new phase angles
are
\begin{equation}\label{theta1p}
\theta'_1 =  \frac{S+1}{2} \theta_1 + \frac{S-1}{2}\theta_2 + \frac{S-1}{2} S\pi
\end{equation}
and
\begin{equation}\label{theta2p}
\theta'_2 = -\frac{S+1}{2} \theta_1 - \frac{S-1}{2}\theta_2 - \frac{S+1}{2} S\pi.
\end{equation}
\vspace{.0 mm}

In the special case of the lowest topological strength $S=1$ of a polar vector field, one obtains
\begin{equation}
\theta'_1 =    \theta_1  ,\quad \theta'_2 = - \theta_1 - \pi,
\end{equation}
whereas for the lowest strength half-integer defects of nonpolar (nematic) orientation fields, $S=1/2$, one finds
\begin{equation}
\theta'_1 =  \frac{3}{4} \theta_1 - \frac{1}{4}\theta_2 - \frac{1}{8} \pi
,\quad
\theta'_2 = -\frac{3}{4} \theta_1 + \frac{1}{4}\theta_2 - \frac{3}{8} \pi.
\end{equation}

One can easily recognize that $\theta'_1+ \theta'_2+ S\pi = 0$ gives the correct boundary condition $\theta'_\infty=0$ in all cases.

The phase mismatch $\delta\theta$ remains invariant under transformations of the coordinates because it is defined by
the reorientation of the c-director along the straight line connecting the two defect cores.
The defect misalignment angle changes to
\begin{equation}
\varphi'_D = \varphi_D -\theta_\infty = \varphi_D- \frac{\theta_1}{2}- \frac{\theta_2}{2}-\frac{S\pi}{2}.
\end{equation}
We replace $\varphi_D$ by the misalignment angle $\delta\theta$ and obtain
\begin{equation}
\varphi'_D = - \frac{\delta\theta}{2S} - \frac{S-1}{2S} \theta_2 - \frac{S+1}{2S} \theta_1- \frac{S+1}{2}\pi.
\end{equation}
In terms of the primed quantities (angles respective to the infinite c-director), using Eqs.~(\ref{theta1p},\ref{theta2p}), one obtains
\begin{equation}
\label{eq:S9}
\varphi'_D = - \frac{\delta\theta}{2S} - \frac{1}{S} \theta'_1 - \pi.
\end{equation}

\begin{figure*}[htbp]
\center
\includegraphics[width=0.73\textwidth]{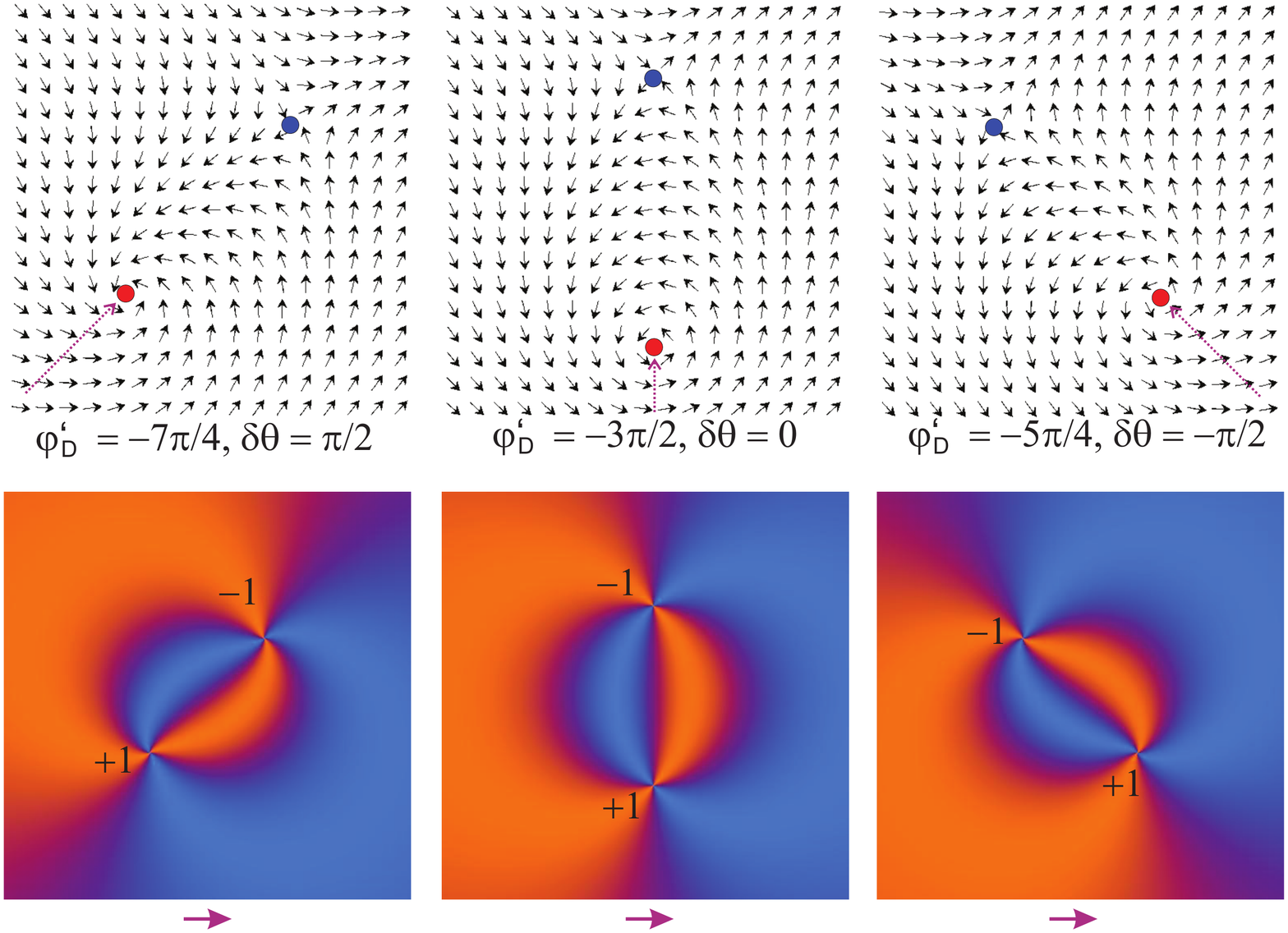}
\caption{Equilibrium c-director orientations around defect pairs with strengths $\pm$1, calculated with Eq.~(\ref{eq:S1}) for fixed defect positions.  $\delta\phi = (-\pi/4,0,\pi/4)$ from left to right. The defect cores are marked by circles  (red for $+S$, blue for $-S$). The top row shows the c-director
fields, with orientation $\theta_\infty=0$ (magenta arrows) in infinity. The bottom row sketches the corresponding optical images with crossed polarizers and diagonally inserted wave plate. Dotted arrows: see text.
}
\label{fig:S1}
\end{figure*}

\begin{figure*}[htbp]
\center
\includegraphics[width=0.73\textwidth]{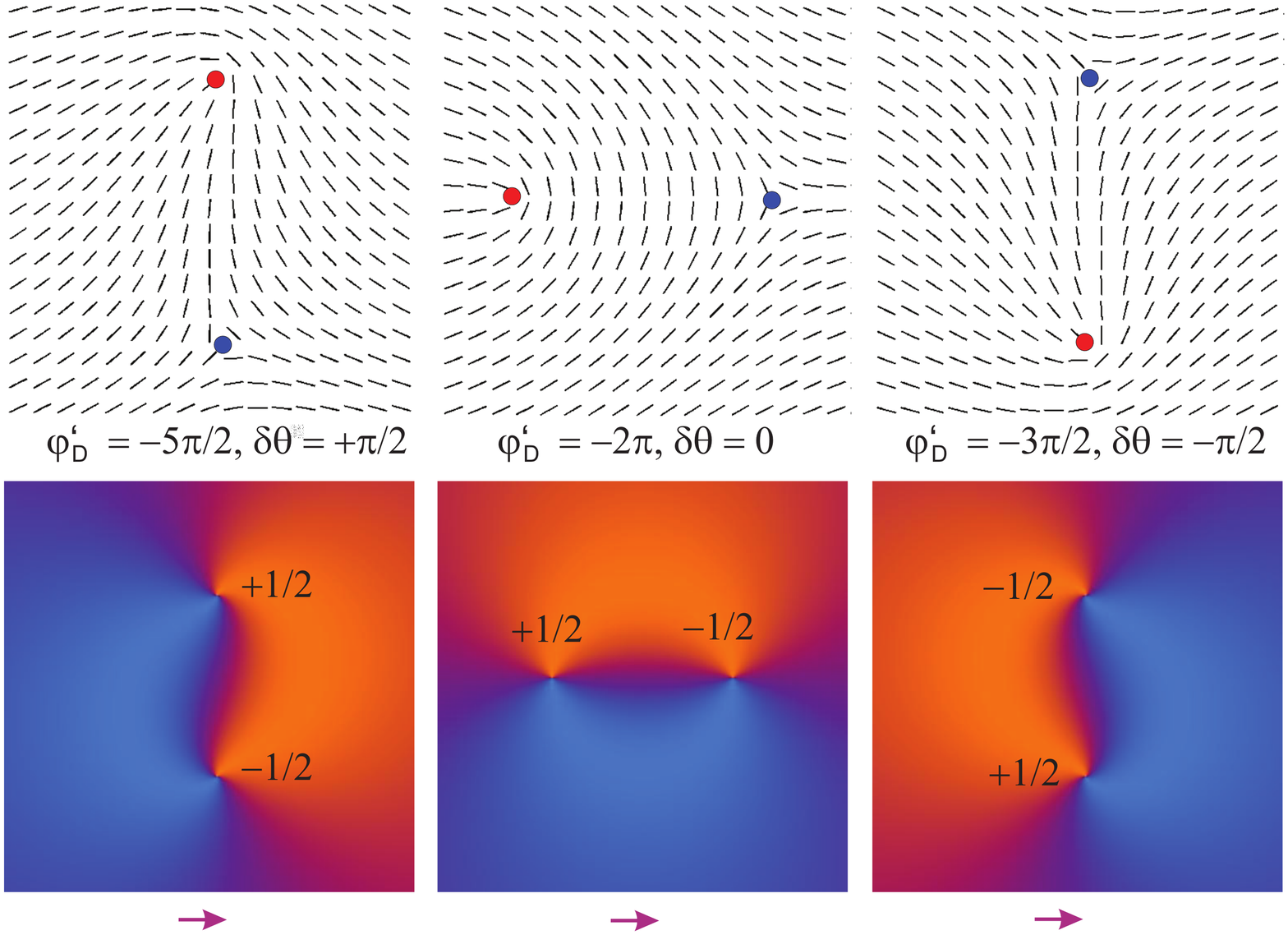}
\caption{Equilibrium director orientations around defect pairs with strengths $\pm 1/2$, calculated with Eq.~(\ref{eq:S1}) for fixed defect positions. The defect cores are marked by circles  (red for $+S$, blue for $-S$). The top row shows the c-director
fields, with orientation $\theta_\infty=0$ (magenta arrows) in infinity. The bottom row sketches the corresponding optical images with crossed polarizers and diagonally inserted wave plate.}
\label{fig:S2}
\end{figure*}

\begin{figure*}[htbp]
\center
\includegraphics[width=0.73\textwidth]{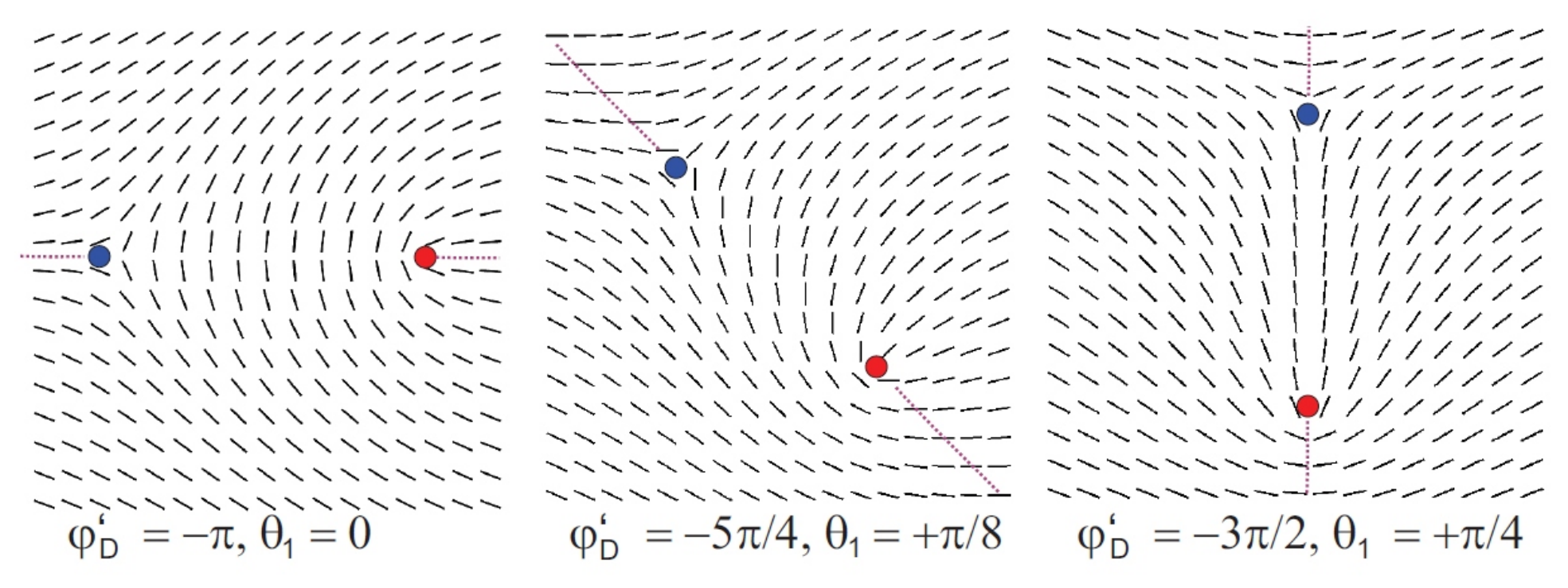}
\caption{Mutually matching director orientations around defect pairs with strengths $\pm 1/2$, calculated with Eq.~(\ref{eq:S1}) for fixed defect positions
and different phase angles $\theta_1$ of the $+1/2$ defect (red dots).
Since $\delta\theta=0$ for all these configurations, they can annihilate without rotating the defects (i.~e. without changing $\theta_{1,2}$).
Dotted lines: see text.}
\label{fig:S3}
\end{figure*}

These solutions are visualized in Figs.~\ref{fig:S1},\ref{fig:S2}. The c-director or director fields, resp., are
shown for three situations in each figure. In the middle, the mutually matching, aligned pairs are shown. In both figures, we have arbitrarily
chosen $\theta_1=\pi/2$. In the $|S|=1$ case, this corresponds to our experimental situation. In general, other orientations may
be conceived as well. In particular, in (strongly polar) smectic C films with $K_{S}<K_{B}$, one could have the situation $\theta_1=0$ or $\theta_1=\pi$.
In the general (nematic) $|S|=1/2$ case, there exist solutions for arbitrary angles $\theta_1$ that are equivalent.
This is shown exemplarily in Fig.~\ref{fig:S3}.

In the special case of $|S|=1$, the geometrical interpretation of the misalignment is immediately evident:
$\varphi_D'=-\pi-\theta_1$ is the angle of the c-director near the core of the positively
charged defect at the side opposite to the conjugate defect (head of dotted arrows in Fig.~\ref{fig:S1}).
This means, when two aligned and mutually matching defects annihilate, they leave a
non-distorted c-director field behind. If $\varphi'_D\neq -\pi-\theta_1$, then the c-director on the connecting axis at both sides of the pair differs from the boundary condition $\theta_\infty'=0$, this is denoted by the term ''misalignment'' ($\delta\phi$) of the pair. After a hypothetical annihilation of such a pair, the c-director field would be defect-free but distorted at the annihilation site, it would have to rotate locally to become uniform.
Thus, it is justified to describe the misalignment by the angle $\delta\phi = \varphi'_D+\pi+\theta_1$. In equilibrium configurations defined by Eq.~(\ref{eq:S1}), for fixed defect positions, this misalignment is linearly related to the mismatch. In the experiments with the free-standing
smectic films, a similar linear relation is found, although with a different factor.

This situation is specific for the $\pm 1$ pair. In all other cases, the situation is more complex
and a simple geometrical interpretation of misalignment is not immediately evident. This is demonstrated in Fig.~\ref{fig:S3}.
The figure shows three examples of $\pm 1/2$ defect pairs that match each other ($\delta\theta=0$). The relation given in
Eq.~(\ref{eq:S9}) is fulfilled
in all three situations, and the defects can annihilate each other approaching on a straight path, leaving an undistorted director
with orientation $\theta_\infty'=0$ behind. Note that this orientation is (and remains) equal to that of the director on the connecting
line, at the sides opposing the partner defect, as indicated by dotted lines in the figure. For the general definition of misalignment of defect
pairs, one may introduce $\delta\phi$ in analogy to the $|S|=1$ case from Eq. ~(\ref{eq:S9}) by
\begin{equation}
\delta\phi=\varphi'_D+ 2\theta_1+\pi
\label{eq:S10}.
\end{equation}
Then, the defect pairs in Fig.~\ref{fig:S2} would have $\delta\phi=(-\pi/2,0,\pi/2)$ and the defect pairs would have $\delta\phi=0$.
The theoretical relation between misalignment and mismatch angles in one-constant approximation would be $\delta\phi=-\delta\theta$.

\section*{Appendix B: Beyond one-constant approximation: Distribution of splay and bend contributions}

\begin{figure*}[htbp]
\center
\includegraphics[width=0.7 \textwidth]{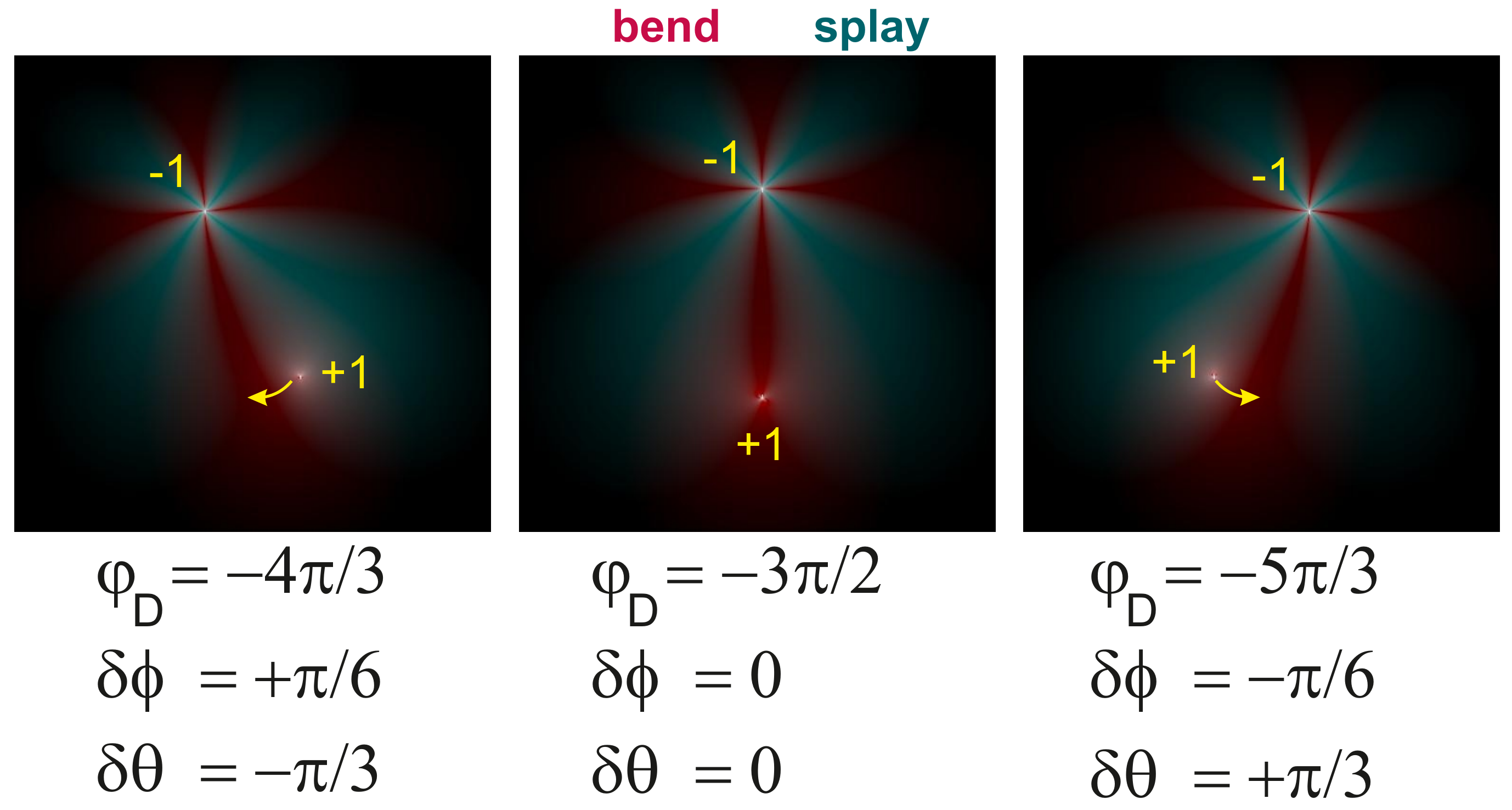}
\caption{Distribution of splay (green) and bend (purple) regions in the $xy$ plane in the vicinity of a defect pair
with different alignment angles $\varphi_D$ in the equilibrium director configurations calculated with Eq.~(\ref{eq:S1}).
The color saturation reflects the logarithm of the splay and bend energy, resp.
In the case of $K_S>K_B$, it is expected that the splayed (green) regions spread out, while the bent (purple) regions contract.
This suggests that the $+1$ defect is relocated closer to the $-3\pi/2$ alignment respective to the $-1$ defect (yellow arrows)
in both misaligned geometries, increasing the discrepancy to the experimentally observed $\delta\phi\approx-\delta\theta$.}
\label{fig:S4}
\end{figure*}

We will analyze here whether the deviation of the measured $\delta\phi(\delta\theta)$ from the theory can be attributed
to the elastic anisotropy present in the experiment, but neglected in the analytical model?
The splay and bend contributions to the elastic deformations of the c-director are not uniformly distributed around the
defect pair. This is irrelevant in one-constant approximation where both contributions are related to an average elastic constant $K$.
In case of $K_B\neq K_S$, the equilibrium configurations are no longer computable from solutions of the Laplace equation. Analytical solutions are only available for very special geometries. In the present problem of the defect pair, the
non-linear differential equation describing the equilibrium has to be solved numerically. The numerical calculation of the exact
energies of the equilibrium solutions in that case is not straightforward because the energy density becomes very large near the
defect cores. We can, however, estimate the influence of an elastic energy by considering the distribution of splay and bend
in the c-director field. Figure \ref{fig:S4} shows three situations where a $\pm1$ defect pair is aligned (center) and misaligned by
a positive or negative $\delta\phi$ (left and right, resp.). The logarithm of the local elastic energy densities of the splay (green) and bend (purple) terms were calculated and color-coded.

In the situation present in our experiment, $K_S=2.2~K_B$, the
system will tend to compress bent regions in favor of expanding splay regions (compared to the one-constant solutions). One may expect that in Fig.~\ref{fig:S4}, the bent c-director region (purple) below the $-1$ defect will contract, the splayed (green) regions at both sides of this region will expand. In a simplistic
interpretation, this means that in the left image, where $\delta\phi$ is positive, the $+1$ defect will be relocated clockwise
around the $-1$ defect, reducing $\varphi_D'$ and $\delta\phi$. In the right image, the $+1$ defect relocates counterclockwise,
thus increasing $\delta\phi$. This effect is opposite to the observation in the experiment where we find $|\delta\phi|$ much larger
than the value $|\delta\theta|/2$ predicted in the one-constant model.

One may conclude from these considerations that the elastic anisotropy has an influence on the relation between misalignment
and mismatch. However, it seems that it is not the primary cause of the observed differences between experiment and model Eq.~(\ref{eq:S1}) since the elastic anisotropy effects apparently increase the discrepancy. An accurate computation of the actual equilibrium configurations at $K_S\neq K_B$ will be needed to verify this relation quantitatively.

	\subsection*{Acknowledgments}
	The German science foundation DFG is acknowledged for support within project STA 425/42-1, HA 8467/2 and HA 8467/3. The German Space Administration DLR is acknowledged for support within grant 50WM1744. P. S. was supported by a DAAD/M\"OB mobility grant.
We thank J. Selinger and X. Tang very much for stimulating discussions.
	%\end{acknowledgments}

%\bibliographystyle{prbst}
\providecommand*{\mcitethebibliography}{\thebibliography}
\csname @ifundefined\endcsname{endmcitethebibliography}
{\let\endmcitethebibliography\endthebibliography}{}

\end{document}